\documentclass[12pt]{iopart}
\def\trade{{\bigcirc}\!\!\!\!\!\mbox{{\tiny R}}}
\def\mathmath{{\it Mathematica}$^{\trade}$\,\,}
\begin{document}

\title[Airy function zeros and quantum mechanics]{Constraints on Airy function zeros
\\ from quantum-mechanical sum rules}

\author{M Belloni$^1$ and R W Robinett$^2$}

\address{$^1$
Physics Department,
Davidson College,
Davidson, NC 28035, USA}

\address{$^2$
Department of Physics,
The Pennsylvania State University,
University Park, PA 16802, USA

\ead{mabelloni@davidson.edu, rick@phys.psu.edu}

}

\begin{abstract}
We derive new constraints on the zeros of Airy functions by using the so-called
quantum bouncer system to evaluate quantum-mechanical sum rules and perform perturbation
theory calculations for the Stark effect. Using
commutation and completeness relations, we show how to systematically evaluate sums
of the form $S_{p}(n) = \sum_{k \neq n} 1/(\zeta_{k}- \zeta_{n})^p$,
for natural $p > 1$, where $-\zeta_{n}$ is the $n^{th}$ zero of $Ai(\zeta)$.
\end{abstract}

\maketitle

\section{Introduction}

The study of special functions and orthogonal polynomials as solutions of
some of the most important and physically relevant model problems in
quantum theory has formed one of the foundations of quantum mechanics since
its earliest days. Familiar  examples include Hermite and Laguerre polynomials for
the harmonic oscillator and Coulomb problems, spherical harmonics (Legendre
polynomials) for angular momentum, Bessel functions (in two and three dimensions) for
both scattering and bound-state problems, Mathieu functions for the quantum-mechanical version of
the pendulum, and even trigonometric functions for the infinite square well problem.

The Airy function \cite{airy_function_monograph} has found many applications
in classical physics, especially in optics and fluid mechanics, and as an important
tool in the derivation of the WKB approximation in quantum mechanics. Airy functions are
also central to the `quantum bouncer' problem, a point mass subject to the potential
\begin{equation}
V(x) =
\left\{
\begin{array}{ll}
Fx     & \mbox{for $x > 0$} \\
\infty & \mbox{for $x \leq 0$}
\end{array}
\right.
\, .
\label{bouncer_potential}
\end{equation}
This problem is amenable
to a variety of approximate treatments (WKB approximation,
variational methods, numerical approaches including the shooting method),
all of which can be compared to the exact
solutions, given in terms of the Airy function $Ai(\zeta)$. This system has
also received considerable recent interest as it is the simplest model for
the {\it ``quantum states of neutrons in the Earth's gravitational field''}
observed by Nesvizhevsky {\it et al.} \cite{neutron_bound_states}.

Gea-Banacloche \cite{bouncing_ball} investigated the time-dependence
of localized wave packets in such a potential, motivated by the
possible application to ultra-cold atoms dropped onto an `atomic mirror,'
and demonstrated the existence of quantum wave packet
revivals \cite{robinett_quantum_revivals} which have been observed in a wide
variety of quantum-mechanical systems. He also found for the first time numerical evidence for
expressions for the normalization of the `quantum bouncer'
energy eigenstates and dipole matrix elements, $\langle n |x| k \rangle$, all of which were
quickly shown to be analytically correct by Vall\'{e}e
\cite{vallee} using earlier published work from the
mathematical \cite{albright} and scientific literature
\cite{gordon} on Airy functions. Goodmanson \cite{goodmanson}
extended these results to produce
closed-form expressions for the
matrix elements of arbitrary powers of position,
$\langle n |x^q| k\rangle$, using
recursion relations. The lowest-order (dipole and quadrupole)
off-diagonal elements, $\langle n|x^{1,2}|k\rangle$, have very simple forms and
were found to depend on inverse powers of the combination
$(\zeta_n - \zeta_k)$ where $-\zeta_{n}$ is the $n^{th}$ zero of $Ai(\zeta)$.

Such differences are intrinsically related to the corresponding differences
in energy eigenvalues for the stationary states, $E_n - E_k$, since
for the `quantum bouncer,' the quantized energy eigenvalues are given
directly by
\begin{equation}
E_n = \zeta_{n} {\cal E}_0
\qquad
\mbox{where}
\qquad
{\cal E}_0 = \left(\frac{\hbar^2 F^2}{2m}\right)^{1/3}
\, .
\end{equation}
The association of such energy differences in conjunction
with dipole (and higher) matrix elements for a quantum-mechanical system
is most familiar from energy-weighted sum rules,
such as the Thomas-Reiche-Kuhn (TRK) formula \cite{trk_sum_rule}
\begin{equation}
\sum_{k} (E_k - E_n) |\langle n |x| k \rangle|^2
= \frac{\hbar^2}{2m}
\qquad
\qquad
\mbox{(TRK sum rule)}
\label{trk_sum_rule}
\, .
\end{equation}
Because of the simple nature of the expressions found in
Refs.~\cite{bouncing_ball} - \cite{goodmanson}, such sum rules automatically
provide constraints on sums of inverse powers of differences of the
zeros of Airy functions, and a systematic study of such relationships
is the topic  which we will examine here. (We note that Sukumar
\cite{sukumar_1}, \cite{sukumar_2} has found constraints on combinations of
inverse powers of energy eigenvalues using Green's function techniques for
several familiar model systems \cite{sukumar_1}, including the linear
potential involving the Airy function \cite{sukumar_2} studied here.
Our approach is different, however, and we find new constraints on the
$\zeta_n$ which have not, to our knowledge, appeared in the mathematical
physics literature.)

In Section~\ref{sec:specific_results} we briefly review the quantum-mechanical
solution of the `quantum bouncer' problem in terms of Airy functions, and use a number of
well-known quantum sum rules to find new constraints on Airy function zeros.
Specifically, we find closed form expressions for the quantity
\begin{equation}
S_{p}(n) \equiv \sum_{k \neq n} \frac{1}{(\zeta_{k} - \zeta_{n})^p}
\label{zero_sums}
\end{equation}
for numerous natural (positive integer) values of $p$. 
It has also been noted that
energy-difference weighted sum rules have a mathematical structure similar
to second-order perturbation theory  \cite{mario_sum_rules}
and we will also use results for the
second-order Stark shift (the energy change due to the addition of an external
constant field) in the same way. In Section~\ref{sec:general_results} we then show
how to systematically derive closed form expressions for the $S_{p}(n)$ in
Eqn.~(\ref{zero_sums}) using  commutation relations and closure methods
(inserting a complete set of states) for all natural values of $p > 1$,
and exhibit specific results for $p = 2,...,11$ as examples. We also discuss,
in Section~\ref{sec:other_constraints},
the relationship of such results to constraints arising from other quantum-mechanical
expressions, such as the famous Bethe sum rule \cite{bethe_sum_rule}
and other interconnections between the $S_{p}(n)$. Finally, in Sec.~\ref{sec:higher_order},
we briefly discuss multi-summation expressions, generalizing Eqn.~(\ref{zero_sums}),
motivated by constraints arising from higher orders of perturbation theory.

\section{The quantum bouncer and sum rule constraints}
\label{sec:specific_results}
The Schr\"{o}dinger equation for the potential in
Eqn.~(\ref{bouncer_potential}) is
\begin{equation}
- \frac{\hbar^2}{2m} \frac{d^2 \psi_{n}(x)}{dx^2} + Fx \psi_n(x)
= E_n \psi_n(x)
\qquad
\qquad
\mbox{for $0 \leq x <\infty$}
\label{eq:se_x}
\end{equation}
where $\psi_n(x)$ and $E_n$ are the energy eigenfunctions and eigenvalues,
respectively. The appropriate boundary conditions are  $\psi_n(x=0) = 0$ and
$\psi_n(x \rightarrow \infty) = 0$.
A change of variable, $x = \rho \zeta$, transforms
Eqn.~(\ref{eq:se_x}) into
\begin{equation}
\frac{d^2 \psi_{n}(\zeta)}{d \zeta^2} = (\zeta - \overline{\zeta}_n)\psi_n(\zeta)
\label{scaled_airy}
\end{equation}
where
\begin{equation}
\rho = \left( \frac{\hbar^2}{2mF}\right)^{1/3}
\qquad
\qquad
\mbox{and}
\qquad
\qquad
E_n \equiv
\left(\frac{\hbar^2 F^2}{2m}\right)^{1/3} \overline{\zeta}_n
= {\cal E}_0 \overline{\zeta}_n
\, .
\end{equation}
The solutions of Eqn.~(\ref{scaled_airy})
are Airy functions, namely $Ai(\zeta - \zeta_n)$ and $Bi(\zeta - \zeta_n)$,
and only the $Ai(\zeta)$ solution is acceptable since the 
$\psi_n(x)$ must be square integrable over the range $(0,\infty)$. 
The boundary condition at the infinite wall ($\zeta=0$) imposes the additional
constraint that $Ai(-\overline{\zeta}_n) = 0$,
so that the $-\overline{\zeta}_n$  are simply the zeros of the
relevant Airy function, $- \zeta_n$.  Since $\overline{\zeta}_n=\zeta_n$,
the energy eigenvalues are then directly given by
$E_n = \zeta_n {\cal E}_0$ and familiar WKB arguments or handbook results
\cite{stegun} can be used to derive an approximate formula for $\zeta_n$
for large $n$ (quantum number), namely
\begin{equation}
\zeta_n \sim \left[ \frac{3\pi}{2}(n-1/4)\right]^{2/3}
\label{large_n_limit}
\, . 
\end{equation}
The normalization of these states, found first
numerically in Ref.~\cite{bouncing_ball},  and then confirmed analytically
in Refs.~\cite{vallee} and \cite{goodmanson}, make use of earlier
published results (Refs.~\cite{albright} and \cite{gordon}), and
can be written in the form
\begin{equation}
\psi_{n}\left(\frac{x}{\rho}\right)
=
\psi_{n}(\zeta) = \frac{Ai(\zeta - \zeta_{n})}{\sqrt{\rho}\, |Ai'(-\zeta_{n})|}
\end{equation}
if we include the proper dimensional constant.

Goodmanson \cite{goodmanson} has found a
recursion relationship
for the position-matrix elements, which when written in terms of the normalized
eigenstates and the scaled variable $\zeta = x/\rho$, reads
\begin{eqnarray}
\!\!\!\!\!\!\!\!\!\!\!\!\!\!\!\!\!\!\!
2 \delta_{1,p} (-1)^{n-k+1} & = &
p(p-1)(p-2)(p-3) \langle n | \zeta^{p-4} | k \rangle
+ 4p(p-1)\zeta_{ave} \langle n |\zeta^{p-2}|k\rangle
\nonumber \\
& &
- 2p(2p-1) \langle n | \zeta^{p-1} | k \rangle
+ (\zeta_n - \zeta_k)^2 \langle n | \zeta^p | k \rangle
\label{recursion_relation}
\end{eqnarray}
where $\zeta_{ave} \equiv (\zeta_n + \zeta_k)/2$. It is understood that
for a given value of $p$, any expectation values of negative powers of
$\zeta$ are to be ignored.

Using this algorithm, one can find the first few diagonal and off-diagonal
matrix elements needed for various sum rule calculations. For example, one
has
\begin{eqnarray}
p = 2: & \qquad & \langle n |\zeta   |n \rangle = \frac{2\zeta_n}{3}
\label{diagonal_1} \\
p = 3: & \qquad & \langle n |\zeta^2 |n \rangle = \frac{8 \zeta_n^2}{15}
\label{diagonal_2} \\
p = 4: & \qquad & \langle n |\zeta^3 |n \rangle = \frac{16\zeta_n^3}{35}
+ \frac{3}{7}
\label{diagonal_3} \\
p = 5: & \qquad & \langle n |\zeta^4|n \rangle =
\frac{128\zeta_n^4}{315} + \frac{80\zeta_n}{63}
\label{diagonal_4} \\
p = 6: &  &
\langle n |\zeta^5|n \rangle =
\frac{256 \zeta_n^5}{693} + \frac{1808\zeta_n^2}{3003}
\label{diagonal_5}
\end{eqnarray}
and
\begin{eqnarray}
\!\!\!\!\!\!
p = 1: & \quad & \langle n |\zeta   | k \rangle= \frac{2\,(-1)^{n-k+1}}{(\zeta_k - \zeta_n)^2}
\label{off_diagonal_1} \\
\!\!\!\!\!\!
p = 2: & \quad & \langle n |\zeta^2 | k \rangle= \frac{24 \,(-1)^{n-k+1}}{(\zeta_k - \zeta_n)^4}
\label{off_diagonal_2} \\
\!\!\!\!\!\!
p = 3: & \quad & \langle n |\zeta^3 | k \rangle = (-1)^{n-k+1}\left [
\frac{720}{(\zeta_k - \zeta_n)^6}
-\frac{48\zeta_n}{(\zeta_k - \zeta_n)^4}
-\frac{24}{(\zeta_k - \zeta_n)^3}
\right ]
\label{off_diagonal_3} \\
\!\!\!\!\!\!
p = 4: & \quad & \langle n |\zeta^4 | k \rangle = (-1)^{n-k+1}\left [
\frac{40340}{(\zeta_k - \zeta_n)^8}
- \frac{3840 \zeta_n}{(\zeta_k - \zeta_n)^6}
- \frac{1920}{(\zeta_k - \zeta_n)^5}
\right ]
\, .
\label{off_diagonal_4}
\end{eqnarray}
(We can then insert the appropriate powers of
$\rho$ as needed for dimensional correctness in physical matrix elements.)

We note that
Eqn.~(\ref{diagonal_1}) is consistent with the quantum-mechanical virial
theorem as it gives
\begin{equation}
\langle n |V(x)|n\rangle = F\rho \langle n |\zeta| n \rangle
= {\cal E}_0  \left(\frac{2\zeta_n}{3}\right)
= \frac{2}{3}E_n
\end{equation}
which is known to be appropriate for a symmetric power-law potential, 
$V_{k}(x) = V_0|x/a|^k$, with $k=1$. Other physically useful matrix elements, 
such as those involving
$\hat{p}^2$, can be obtained by writing $\hat{p}^2 = 2m(\hat{H} - Fx)$. For
example, this allows for the evaluation of diagonal matrix elements such as
\begin{equation}
\langle n | \hat{p}^2 | n \rangle =
2m {\cal E}_{0}\left(\frac{\zeta_n}{3}\right)
\qquad
\mbox{and}
\qquad
\langle n | \hat{p}^4 | n \rangle =
(2m{\cal E}_0)^2 \left(\frac{\zeta_n^2}{5}\right)
\label{momentum_squared_ones}
\end{equation}
or off-diagonal ones such as
\begin{equation}
\langle n | \hat{p}^2 |k \rangle
= -2mF\rho \langle n |\zeta| k \rangle
= -\frac{4{\cal E}_0\,(-1)^{n-k+1}}{(\zeta_k - \zeta_n)^2}
\,.
\label{offdiagonal_momentum}
\end{equation}

Using these results, we can begin cataloging the constraints which result
from the application of various well-known quantum-mechanical sum rules.
For example, using the energy differences
$E_k-E_n = {\cal E}_0 (\zeta_k - \zeta_n)$ and the off-diagonal
dipole matrix elements in Eqn.~(\ref{off_diagonal_1}),
we start with the most-cited sum rule, the one formulated by
Thomas-Reiche-Kuhn as in Eqn.~(\ref{trk_sum_rule}).
We find that all of the dimensional parameters cancel,
leaving the simplest constraint we encounter, namely
\begin{equation}
S_{3}(n) \equiv \sum_{k \neq n} \frac{1}{(\zeta_k - \zeta_n)^3}
= \frac{1}{4}
\, .
\end{equation}
This example and all of the other constraint equations for the $S_{p}(n)$
 derived here, as well as the on-diagonal and off-diagonal matrix elements
in Eqns.~(\ref{diagonal_1}) - (\ref{off_diagonal_4}),
can be verified to essentially arbitrary accuracy by
using the numerical ability of mathematical manipulation programs
such as \mathmath.

The so-called `monopole sum rule,'  which has been used in applications
to nuclear collective excitations \cite{bohigas}, is given by
\begin{equation}
\sum_{k} (E_k - E_n)|\langle n |x^2|k\rangle|^2 =
\frac{2\hbar^2}{m} \langle n |x^2|n \rangle
\, ,
\label{monopole_sum_rule}
\end{equation}
and is of a similar form to the TRK sum rule.
 Using the $\zeta^2$ matrix elements from
Eqns.~(\ref{diagonal_2}) and (\ref{off_diagonal_2}), we find the constraint
equation
\begin{equation}
S_{7}(n) \equiv \sum_{k \neq n} \frac{1}{(\zeta_k - \zeta_n)^7}
= \frac{\zeta_n^2}{270}
\, .
\end{equation}
We note that the  monopole sum rule in Eqn.~(\ref{monopole_sum_rule})
is a special case of a more general form derived by
Wang \cite{wang_sum_rule},
\begin{equation}
\sum_{k} (E_k - E_n) |\langle n|F(x)|k\rangle|^2
= \frac{\hbar^2}{2m}
\left\langle
n \left|
\frac{dF(x)}{dx}
\,
\frac{dF^{*}(x)}{dx}
\right| n \right\rangle
\end{equation}
which simplifies if the function is real so that $F(x) = F^{*}(x)$.
This general result can also be used to immediately reproduce the TRK sum
rule by using $F(x) = x$.

Bethe and Jackiw \cite{bethe_intermediate,jackiw_sum_rules}
derive several other sum rules for dipole moment matrix elements
by using multiple commutation relations with the Hamiltonian,
thus yielding higher powers of the energy difference.
These higher-order sum rules include:

\begin{equation}
\sum_{k} (E_k - E_n)^2 |\langle n |x|k \rangle|^2
=\frac{\hbar^2}{m^2} \langle n |\hat{p}^2|n\rangle
=\frac{2\hbar^2}{m} \left [ E_n - \langle n|V(x)|n\rangle \right ]
\label{second_power_momentum_sum_rule}
\end{equation}
\begin{equation}
\sum_{k}(E_k - E_n)^3 |\langle n |x| k \rangle|^2
= \frac{\hbar^4}{2m^2} \left\langle n \left|
\frac{d^2 V(x)}{dx^2} \right| n \right\rangle
\label{first_potential_sum_rule}
\end{equation}
and
\begin{equation}
\sum_{k} (E_k - E_n)^4 |\langle n |x| k \rangle|^2
= \frac{\hbar^4}{m^2} \left\langle n \left|\left(\frac{dV(x)}{dx}\right)^2
\right| n \right\rangle
\label{second_potential_sum_rule}
\end{equation}
where Eqns.~(\ref{first_potential_sum_rule}) and (\ref{second_potential_sum_rule}) are sometimes called the
``{\it force times momentum}'' and ``{\it force squared}'' sum rules,
respectively. We recall that not all such sum rules are guaranteed to lead
to convergent expressions.

The first of these three higher-order sum rules,
Eqn.~(\ref{second_power_momentum_sum_rule}), gives the relation
\begin{equation}
S_{2} = \sum_{k \neq n} \frac{1}{(\zeta_k - \zeta_n)^2}
= \frac{\zeta_n}{3}
\,.
\end{equation}
This sum rule leads to a convergent result since, for large $k$, the
terms in the summation scale as $1/k^{4/3}$. The sum rule results in
Eqns.~(\ref{first_potential_sum_rule}) and (\ref{second_potential_sum_rule}),
however, do not converge due to the derivatives of the discontinuous potential
energy function in Eqn.~(\ref{bouncer_potential}). This implies that $S_{1}(n)$
is not convergent.

As mentioned above, it has been emphasized that the standard expression
for the second-order energy shift in perturbation theory due to an
added potential energy term of the form $\overline{V}(x)$,  given by
\begin{equation}
E_n^{(2)} = \sum_{k \neq n}
\frac{|\langle n |\overline{V}(x)|k\rangle|^2}{E_{n}^{(0)} - E_{k}^{(0)}}
\label{second_order_shift}
\, ,
\end{equation}
is also a form of energy-weighted sum rule. The authors of
Ref.~\cite{mario_sum_rules} have used this fact to evaluate the Stark
shift, that is the second order energy shift due to an external constant
field, with a potential of the form $\overline{V}(x) = \overline{F}x$, in
two model systems, the infinite square well and single attractive
$\delta$-function potential. Then, using the same mathematical techniques
as for the confirmation of many other sum rules in those two cases (where the
relevant tools are the Mittag-Lefler theorem and standard contour integration
methods respectively) one can evaluate $E_n^{(2)}$ in closed form.

In this situation, the addition of a uniform external field $\overline{V}(x)
= \overline{F}x$ to the potential in Eqn.~(\ref{bouncer_potential}) leads
to a soluble problem with a simple redefinition of the constant force,
$F \rightarrow F + \overline{F}$, giving an exact value for the new
energy eigenvalues,
\begin{equation}
\!\!\!\!\!\!
\tilde{E}_n = \zeta_n \left[\frac{\hbar^2(F+ \overline{F})^2}{2m}\right]^{1/3}
= E_n \left[1 + \frac{\overline{F}}{F}\right]^{2/3}
\qquad
\mbox{where}
\qquad
E_n = {\cal E}_0 \zeta_n
\, .
\label{exact_result}
\end{equation}
The term in brackets can be easily expanded giving predictions for the
first-, second-, and third-order perturbation theory results, namely
\begin{equation}
\!\!\!\!\!\!\!\!\!\!\!\!\!\!\!\!\!\!\!\!\!\!\!\!\!\!\!\!
E_n^{(1)} =  \frac{2}{3}\left(\frac{\overline{F}}{F}\right)
\left( {\cal E}_0 \zeta_{n} \right)\,,
\quad
E_n^{(2)}  =  -\frac{1}{9}\left(\frac{\overline{F}}{F}\right)^2
\left( {\cal E}_0 \zeta_{n} \right)\,,
\quad
E_n^{(3)} = \frac{4}{81}\left(\frac{\overline{F}}{F}\right)^3
\left( {\cal E}_0 \zeta_{n} \right)
\label{third_order_stark}
\, .
\end{equation}
The first-order result is easily confirmed by noting that
\begin{equation}
E_n^{(1)} = \langle n |\overline{F}x|n \rangle
= \overline{F}\rho \langle n|\zeta|n\rangle
= \left(\frac{\overline{F}}{F}\right)(F\rho) \left(\frac{2 \zeta_n}{3}\right)
= \frac{2}{3}\left(\frac{\overline{F}}{F}\right) ({\cal E}_0 \zeta_n)
\end{equation}
where we use the diagonal dipole matrix element in Eqn.~(\ref{diagonal_1}).
The second-order shift equation can then be used as a new
constraint on a different combination of Airy function zeros, giving
\begin{equation}
S_{5}(n) = \sum_{k\neq n}\frac{1}{(\zeta_k - \zeta_n)^5}
= \frac{\zeta_n}{36}
\end{equation}
and we note that $S_{2}(n) = 12 S_{5}(n)$. Higher-order perturbative
corrections can in principle be used to derive closed form expressions
for more complex combinations of inverse powers of $(\zeta_k - \zeta_n)$,
since one can expand the exact result in Eqn.~(\ref{exact_result})
to arbitrarily high order.  In fact, we use the third-order expression
to briefly discuss multi-index summation
generalizations of Eqn.~(\ref{zero_sums}) in Sec.~\ref{sec:higher_order}.

To briefly summarize, the evaluation of
several well-known quantum-mechanical sum rules and the evaluation of the
related energy-weighted sum over dipole matrix elements from perturbation
theory have provided closed form expressions for
$S_{p}(n)$ for $p = 2, 3, 5$ and $7$. In the next section, we show how
to systematically evaluate $S_{p}(n)$ for all natural $p>1$.

\section{Systematic method for construction of $S_p(n)$}
\label{sec:general_results}

Before proceeding, we recall the  techniques that are
used in the derivation of many of the familiar quantum-mechanical sum rules,
especially those involving dipole matrix elements. As an example, for the
TRK sum rule in Eqn.~(\ref{trk_sum_rule}), we write the standard
$x,\hat{p}$ commutation relation, $[\hat{p},x] = -i\hbar$, bracketed by energy eigenstates:
\begin{equation}
-i \hbar = \langle n| \hat{p}x - x\hat{p}|n\rangle
=
\sum_{all\, k} \left\{
\langle n |\hat{p}|k\rangle
\langle k|x|n\rangle
-
\langle n |x |k\rangle
\langle k |\hat{p} |n\rangle
\right\}
\label{basic_one}
\end{equation}
where we have also inserted a complete set of states
to obtain the right-hand side of the equality.
 We can then use a second commutation relation, namely
\begin{equation}
[\,\hat{H},\,x\,]
= \frac{1}{2m}[\,\hat{p}^2,\,x\,]
= \frac{\hbar}{mi} \, \hat{p}
\, ,
\label{two_commutators}
\end{equation}
where we assume a standard one-dimensional Hamiltonian of the form
$\hat{H} = \hat{p}^2/2m + V(x)$,
to write
\begin{equation}
\langle n |\hat{p}|k\rangle
= \frac{im}{\hbar} \langle n|[\,\hat{H},\,x\,]|k\rangle
= \frac{im(E_n-E_k)}{\hbar}
\langle n|x|k\rangle\;.
\label{matrix_element_connection}
\end{equation}
There is a similar expression for $\langle k |\hat{p}|n\rangle$ and
combining these two results in Eqn.~(\ref{basic_one}) gives the TRK sum rule. Note that Eqn.~(\ref{matrix_element_connection}) gives $\langle n|\hat{p}|k \rangle = 0$ if $n =k$ as is appropriate for energy eigenstates where the
average momentum should vanish in a stationary state.

Specializing now to the case of
the quantum bouncer, and using the result in Eqn.~(\ref{off_diagonal_1}),
we find the more specific result,
\begin{equation}
\langle n |\hat{p}|k \rangle = \frac{\hbar}{i\rho}
\frac{(-1)^{n-k+1}}{(\zeta_k - \zeta_n)}
\label{momentum_matrix_element}
\end{equation}
and we note that $\langle n |\hat{p}|k \rangle^*
= \langle k |\hat{p}|n \rangle $.
This is the lowest-order (in inverse powers of $\zeta_k - \zeta_n$)
term possible, and is our starting point.

We see that the important ingredients are insertion of a
complete set of states, the matrix-element connection in
Eqn.~(\ref{matrix_element_connection}), and appropriate commutation
relations. Motivated by these methods, we start with the simple
closure relationship for the momentum operator, namely
\begin{equation}
\sum_{all \, k} \langle n|\hat{p}|k \rangle \langle k |\hat{p} | n \rangle
= \langle n |\hat{p}^2|n \rangle
\,.
\label{momentum_closure}
\end{equation}
The only non-zero elements on the left-hand side are the off-diagonal
matrix elements in Eqn.~(\ref{momentum_matrix_element}), while the
diagonal matrix element on the right-hand side can be evaluated using
Eqn~(\ref{momentum_squared_ones}).
Inserting these results gives
\begin{equation}
S_{2}(n) = \frac{\zeta_n}{3}
\,.
\end{equation}
This is the same result as obtained from the sum rule in
Eqn.~(\ref{second_power_momentum_sum_rule}),
which is correct since that expression is most simply obtained from
the closure relationship in Eqn.~(\ref{momentum_closure}), using
the expression in Eqn.~(\ref{matrix_element_connection}) twice.

The next higher power of ($\zeta_k - \zeta_n)^{-1}$ is obtained by
inserting a complete of states into the commutation relation
$[x,\hat{p}] = i\hbar$, namely
\begin{equation}
\sum_{all\, k}
\left\{
\langle n |x|k \rangle \langle k |\hat{p}| n \rangle
-
\langle n |\hat{p}|k \rangle \langle k |x|n \rangle
\right\}
= \langle n |i\hbar|n \rangle
= i\hbar
\end{equation}
which, of course, reproduces the TRK sum rule result, giving
$S_{3}(n) = 1/4$.  We note here that in any
summation involving $\langle n |\hat{p}|k\rangle$, the $n=k$ term is not
present.

To evaluate $S_{4}(n)$, we use the  $x$-closure relationship,
\begin{equation}
\sum_{all \,k}\langle n |x|k\rangle \langle k |x|n \rangle = 
\label{x_closure_relationship}
\langle n|x^2|n\rangle
\end{equation}
and by explicitly including both on- and off-diagonal terms on the
left-hand side (and removing all dimensional constants) this gives us
\begin{equation}
|\langle n |\zeta |n\rangle|^2 + \sum_{k\neq n}\frac{4}{(\zeta_k - \zeta_n)^4}
= \langle n | \zeta^2|n\rangle
\end{equation}
or
\begin{equation}
\left(\frac{2\zeta_n}{3}\right)^2 + 4 S_{4}(n) = \frac{8\zeta_n^2}{15}
\qquad
\quad
\mbox{yielding}
\quad
\qquad
S_{4}(n) = \frac{\zeta_n^2}{45}
\,.
\end{equation}

The correct iterative procedure required to evaluate $S_{p}(n)$ for
any natural value of $p$ is now clear.  The off-diagonal
matrix elements for $\langle n |\zeta^q |k \rangle$ will have a leading
term of order $(\zeta_k - \zeta_n)^{-2q}$. We then apply closure to
the general commutator result $[x^q, \hat{p}]  = iq\hbar x^{q-1}$ in
the form
\begin{equation}
\sum_{all\, k}
\left\{
\langle n |x^q|k \rangle \langle k |\hat{p} |n \rangle
-
\langle n|\hat{p}|k \rangle \langle k |x^q|n \rangle
\right\}
= iq\hbar \langle n |x^{q-1}|n \rangle
\,.
\end{equation}
The recursive relation in  Eqn.~(\ref{recursion_relation})
can then be used straightforwardly (if tediously)
to evaluate both $\langle n |x^q| k \rangle$ and $\langle n |x^{q-1}|n\rangle$
to obtain a closed-form expression  for $S_{2q+1}(n)$ in terms of explicit
powers of $\zeta_n$ and values of $S_{p}(n)$ for $p<2q$. This procedures allows
one to increment the value of $p$ by one, since the inclusion of the
$\langle n |\hat{p}|k\rangle$ matrix elements adds one more inverse power of
$(\zeta_k - \zeta_n)$.

To evaluate $S_{p}(n)$ for $p$ values two units higher, 
we generalize the $x$-closure relationship in
Eqn.~(\ref{x_closure_relationship}) to
\begin{equation}
\sum_{all\, k} \langle n | x^{q}| k \rangle \langle k|x|n\rangle
= \langle n |x^{q+1}| n\rangle
\end{equation}
and again evaluation of only $x$-dependent on- and off-diagonal matrix
elements will suffice to obtain an expression for $S_{2q+2}(n)$,
as the
$\langle k |x|n\rangle$ term adds two inverse powers of $(\zeta_k - \zeta_n)$.

In this way, we are able to systematically evaluate $S_{p}(n)$ for all
$p>1$, since the $p=1$ case,  corresponding to use of
Eqn.~(\ref{first_potential_sum_rule}),
does not converge. We present below results obtained in this way for the cases
$p=2,...,11$,
\begin{eqnarray}
S_{2}(n) & = & \frac{\zeta_n}{3} 
\label{2result}\\
S_{3}(n) & = & \frac{1}{4} \\
S_{4}(n) & = & \frac{\zeta_n^2}{45} \\
S_{5}(n) & = & \frac{\zeta_n}{36} \\
S_{6}(n) & = & \frac{2\zeta_n^3}{945} + \frac{1}{112} \\
S_{7}(n) & = & \frac{\zeta_n^2}{270} \\
S_{8}(n) & = & \frac{\zeta_n^4}{4725} + \frac{5\zeta_n}{2268} \\
S_{9}(n) & = & \frac{\zeta_n^3}{2100} + \frac{1}{2240} \\
S_{10}(n) & = & \frac{2\zeta_n^5}{93555} + \frac{611\zeta_n^2}{1496880} \\
S_{11}(n) & = & \frac{\zeta_n^4}{17010} + \frac{43\zeta_n}{272160}
\label{11result}
\, ,
\end{eqnarray}
all of which have also been verified numerically.

\section{Other sum rule constraints}
\label{sec:other_constraints}

There are a large number of other possible constraints, obtainable
from both the use of the systematic approach developed above, or through
other familiar sum rules.

For increasing values of $p$, there are often several closure-motivated
constraints that can be applied to obtain closed-form expressions for
$S_{p}(n)$. For example, to evaluate $S_{8}(n)$, we can use either
\begin{equation}
\sum_{all\, k} \langle n | x^3 |k\rangle \langle k |x|n \rangle
= \langle n |x^4 |n \rangle
= \sum_{all\, k}\langle n |x^2|k \rangle \langle k |x^2|n \rangle
\label{first_double_closure}
\end{equation}
and use one as a cross-check against the other. Alternatively, such
identities can be used as further constraints among the various
$S_{p}(n)$. The expression above, for example, requires that
\begin{eqnarray}
&
\left(\frac{16\zeta_n^3}{35} + \frac{3}{7}\right)
\left(\frac{2 \zeta_n}{3}\right)
+ 1440 S_{8}(n) - 96\zeta_n S_{6}(n) - 47 S_{5}(n)
&  \nonumber \\
& = \left(\frac{8 \zeta_n^2}{15}\right)^2
+ 576 S_{8}(n) &
\end{eqnarray}
which is easily verified.

One of the more famous constraints on dipole matrix elements
 is the Bethe sum rule
\cite{bethe_sum_rule}
\begin{equation}
\sum_{k}(E_k-E_n)
|\langle n |e^{iqx}|k\rangle|^2 = \frac{\hbar^2 q^2}{2m}
\, ,
\label{bethe_sum_rule}
\end{equation}
which was developed in the study of energy loss mechanisms,
eventually leading to the Bethe-Bloch formula. This single sum rule
actually provides an infinite tower of constraints on the $S_{p}(n)$.
To see this,  we expand the matrix element of the exponentials via
\begin{equation}
\!\!\!\!\!\!\!\!\!\!\!\!\!\!\!\!\!\!\!\!
\langle n |e^{iqx}| k \rangle
= \langle n |k\rangle
+ iq \langle n |x|k \rangle
- \frac{q^2}{2!} \langle n | x^2| k\rangle
- i\frac{q^3}{3!} \langle n |x^3| k \rangle
+ \frac{q^4}{4!} \langle n |x^4| k \rangle
+ \cdots
\, .
\end{equation}
The odd-order (in $q$) terms on the left-hand side of
Eqn.~(\ref{bethe_sum_rule}) automatically cancel since it is an
even function by construction,
while the ${\cal O}(q^0)$ terms is absent since $\langle n |k \rangle = 0$
if $n\neq k$.  The lowest non-vanishing
term, the ${\cal O}(q^2)$ term on the left-hand side,
saturates the right-hand side by reproducing the TRK sum rule.
All higher-order terms must therefore vanish.
For example, the vanishing of the ${\cal O}(q^4)$ term implies that
\begin{equation}
\frac{1}{4} \sum_{k}(E_k-E_n) \langle n |x^2|k \rangle \langle k |x^2|n \rangle
= \frac{1}{3} \sum_{k}(E_k-E_n) \langle n |x^3|k \rangle \langle k |x|n \rangle
\end{equation}
which is a different constraint equation involving $S_{8}(n)$ and
including lower-order
terms than the one in Eqn.~(\ref{first_double_closure}).

Closure relationships involving mixed combinations of $x$ and $\hat{p}$
of higher order in the momentum operator are also possible, often
reproducing earlier results. For example, the closure-relation for
$\hat{p}^4$, namely
\begin{equation}
\sum_{all\, k} \langle n |\hat{p}^2 | k \rangle
         \langle k | \hat{p}^2| n \rangle
= \langle n | \hat{p}^4 | n \rangle
\end{equation}
can be used in conjunction with Eqns.~(\ref{momentum_squared_ones})
and (\ref{offdiagonal_momentum}) to evaluate $S_{4}(n)$. Wang
\cite{wang_sum_rule} has also derived a number of sum rules involving
mixed position- and momentum-matrix elements which can be used.

\section{Multi-index summations}
\label{sec:higher_order}

To explore more complex relationships involving inverse
powers of $(\zeta_k - \zeta_n)$, we extend the perturbation
theory analysis of the Stark effect for the quantum bouncer to
third order. We first recall that
the third-order correction in perturbation theory due to a
general $\overline{V}(x)$ term is given by
\begin{eqnarray}
E_{n}^{(3)}
& = &
\sum_{k \neq n} \sum_{j \neq n}
\frac{
\langle n |\overline{V}(x)| k \rangle
\langle k |\overline{V}(x)| j \rangle
\langle j |\overline{V}(x)| n \rangle}
{
(E_n^{(0)} - E_{k}^{(0)})
(E_n^{(0)} - E_{j}^{(0)})
}
\nonumber \\
\qquad & &  - \langle n|\overline{V}(x)|n \rangle
\sum_{k \neq n}
\frac{|\langle n |\overline{V}(x)|k\rangle|^2}{(E_{n}^{(0)} - E_{k}^{(0)})^2}
\, .
\end{eqnarray}
Using $\overline{V}= \overline{F}x$ as above,
the exact third-order result in Eqn.~(\ref{third_order_stark}),
and removing all dimensional factors gives the constraint
\begin{eqnarray}
\frac{4\zeta_n}{81}
& = & \sum_{k\neq j \neq n}
\frac{
\langle n |\zeta| k \rangle
\langle k |\zeta| j \rangle
\langle j |\zeta| n \rangle
}
{(\zeta_k - \zeta_n)(\zeta_j - \zeta_n)}
+ \sum_{j=k\neq n}
\frac{|\langle n |\zeta|k\rangle|^2 \langle k |\zeta| k \rangle}
{(\zeta_k - \zeta_n)^2}
\nonumber \\
& &  \qquad
- \langle n|\zeta|n\rangle
\sum_{k \neq n} \frac{|\langle n |\zeta|k\rangle|^2}{(\zeta_k - \zeta_n)^2}
\,.
\end{eqnarray}
In evaluating the double sum, we have separated off the cases where
$j=k \neq n$, leaving the distinct $j\neq k \neq n$ terms.

Inserting the appropriate diagonal and off-diagonal matrix elements in
Eqns.~(\ref{diagonal_1}) and (\ref{off_diagonal_1}), we then find the
relationship
\begin{equation}
\!\!\!\!\!\!\!\!\!\!\!\!\!\!\!\!\!\!\!\!\!\!\!\!\!\!\!\!\!\!\!\!\!\!\!\!\!
-8 \sum_{k \neq j \neq n}
\frac{1}{(\zeta_k - \zeta_n)^3(\zeta_k - \zeta_j)^2(\zeta_j - \zeta_n)^3}
 + \frac{8}{3}\left[ S_{5}(n) + \zeta_n S_{6}(n)\right]
 - \frac{8\zeta_n}{3} S_6(n) = \frac{4\zeta_n}{81}
\end{equation}
or
\begin{equation}
T_{3,2,3}(n) \equiv
\sum_{k \neq j \neq n}
\frac{1}{(\zeta_k - \zeta_n)^3 (\zeta_k - \zeta_j)^2(\zeta_j - \zeta_n)^3}
= \frac{\zeta_n}{324}
\, ,
\end{equation}
a remarkably simple identity, which we have also confirmed numerically.

The presence of double (or higher) summations can be systematized,
as in Sec.~\ref{sec:general_results}, by introducing more than one
insertion of a complete set of states into any quantum identity, either
resulting from simple closure or commutation relations. For example, we
can write
\begin{equation}
\sum_{all\, k} \sum_{all\, j}
\langle n |x|k \rangle
\langle k |x|j \rangle
\langle j |x|n \rangle
= \langle n|x^3|n\rangle
\,.
\end{equation}
In evaluating the double sum, we must consider the following special
cases, namely (i) $n=k=j$, (ii) two equal contributions arising
from $n=k\neq j$ and $n=j \neq k$, (iii) $j=k \neq n$, and (iv) the
completely distinct double sum where $j \neq k \neq n$. Including
all of these possibilities gives the constraint
\begin{eqnarray}
\left(\frac{2\zeta_n}{3}\right)^3
+ 2\left(\frac{8\zeta_n}{3} S_4(n)\right)
+ \frac{8}{3}\left[ S_3(n) + \zeta_n S_4(n) \right]
\qquad \qquad \qquad & & \nonumber\\
- 8 \sum_{j\neq k \neq n}
\frac{1}{(\zeta_k-\zeta_n)^2 (\zeta_j-\zeta_k)^2 (\zeta_j - \zeta_n)^2}
= \frac{16\zeta_n^3}{35} + \frac{3}{7} & &
\end{eqnarray}
or
\begin{equation}
T_{2,2,2} = \frac{2\zeta_n^3}{945} + \frac{5}{168}
\end{equation}
which we have also confirmed numerically. Clearly an infinite number of
multi-index summations can be evaluated in closed form in this way.

\section{Conclusions and future directions}

In conclusion, motivated by identities derived using a variety of
quantum-mechanical sum rules, we have developed techniques to
systematically evaluate sums of the form in Eqn.~(\ref{zero_sums})
for arbitrary natural $p>1$, explicitly exhibiting results for
$S_{p}(n)$ for $p=2,...,11$.  In addition, we have identified many other
additional constraints on the $S_{p}(n)$ arising from
self-consistency of the method and the Bethe sum rule. Using higher-order
perturbation theory results as a starting point, we have also discussed the
existence of multi-summation constraints arising from the the repeated
use of insertion of a complete set of states. Future work might involve
attempts at inductively generated closed form expressions for the
$S_p(n)$ as well as exploration of the algebraic structures suggested 
by the pattern of results in Eqns.~(\ref{2result}) - (\ref{11result}),
where each term only contains powers of $\zeta_n$ modulo three.

\end{document}